\documentclass[prd,nofootinbib,showpacs,preprint]{revtex4}

\usepackage{array} 
\usepackage{amsmath,amssymb} 
\usepackage{graphicx} 
\usepackage{multirow}

\def\vev#1{\langle #1 \rangle}
\def\be{\begin{equation}}
\def\ee{\end{equation}}
\def\bea{\begin{eqnarray}}
\def\eea{\end{eqnarray}}

\def\sss{\scriptscriptstyle}
\def\Ls{{\sss L}}

{
{

\begin{document}

 \title{PeV scale Left-Right symmetry and baryon 
asymmetry of the Universe}
 \author{Anjishnu Sarkar} 
 \email{anjishnu@phy.iitb.ac.in}
 \author{Abhishek} 
 \email{abhishek.aero@iitb.ac.in}
 \author{Urjit A. Yajnik}
 \email{yajnik@phy.iitb.ac.in}
 \affiliation{Indian Institute of Technology, Bombay, Mumbai - 400076, India}
 \date{}




\begin{abstract}
We study the cosmology of two versions of supersymmetric Left-Right symmetric 
model. The  scale of the $B-L$ symmetry breaking in these models is
naturally low, $10^4 - 10^6$ GeV. Spontaneous breakdown of parity 
is accompanied by a first order phase transition. We simulate the domain walls
of the phase transition and show that they provide requisite conditions,
specifically, $CP$ violating phase needed for leptogenesis. Additionally soft 
resonant leptogenesis is conditionally viable in the two models considered. 
Some of the parameters in the soft supersymmetry breaking 
terms are shown to be constrained from these considerations.
It is argued that the models may be testable in upcoming collider and cosmology 
experiments. 
\end{abstract}

\maketitle


\section{Introduction} 
\label{sec:intro} 

Left-Right symmetric model 
\cite{ Pati:1974yy, Mohapatra:1974gc, Senjanovic:1975rk, Mohapatra:1980qe, Deshpande:1990ip}
is a simple extension of the Standard Model (SM) 
\cite{Glashow:1961tr, Weinberg:1967tq, Salam:1968rm}.
From a theoretical point 
of view it provides an elegant explanation for the conservation of 
$B-L$ which automatically becomes a gauge charge, and as a bonus 
provides a natural explanation  for the meaning of the electroweak 
hypercharge. The new gauge symmetries 
required constitute the group $SU(2)_R\otimes U(1)_{B-L}$. The model
has long been understood as a possible intermediate state in the $SO(10)$
\cite{Georgi:1974my, Fritzsch:1974nn} grand unified theory (GUT). 
However unification in $SO(10)$ generically also forces
the possible intermediate scale of Left-Right symmetry to be high
and therefore inaccessible to accelerators. On the other hand, the less 
restrictive principle of exact Left-Right symmetry is still appealing 
though it leaves the $U(1)_{B-L}$ charge unrelated to the two identical 
charges of $SU(2)_L$ and $SU(2)_R$. 
As for the fermion sector the presence of right handed neutrino 
states in the theory allows the possibility of explaining the 
smallness of  the observed neutrino masses\cite{Fukuda:2001nk, Ahmad:2002jz, Ahmad:2002ka, Bahcall:2004mz} from the see-saw mechanism 
\cite{Minkowski:1977sc, Gell-Mann:1980vs, Yanagida:1979as, Mohapatra:1979ia}.  While the scale of Majorana masses is no 
longer as high as in the conventional see-saw expectations, the
PeV scale still  permits \cite{Sahu:2004sb} explaining the smallness
of the light neutrino mass scale for at least certain textures of 
fermion mass parameters.
It is therefore worth exploring the possibility that the scale of 
Left-Right symmetry be the PeV scale, potentially testable in colliders. 

Whether we follow the GUT proposal or the PeV scale possibility, 
the large hierarchy between the mass scales $M_{EW}\sim 250$GeV 
of electroweak symmetry and $M_{GUT}\sim 10^{15}$GeV is difficult  to 
understand within the Higgs paradigm. While the Higgs sector of the 
Standard Model is poorly understood, it is nevertheless very successful.
We therefore speculate that the breaking of both the $SU(2)_L$ and 
$SU(2)_R$ being at a comparable scale will have a similar explanation,
possibly a comprehensive one including both. 
There remains the need to understand  the hierarchy with respect 
to a larger mass scale either the 
GUT scale or the Planck scale. In this 
paper we assume  supersymmetry (SUSY) to be the mechanism
to stabilize the hierarchy beyond the electroweak scale \cite{Witten:1981nf, Kaul:1981hi}, 
in other words we assume TeV scale SUSY
\footnote{See for instance \cite{Martin:1997ns, Aitchison:2005cf, Chung:2003fi}
and references therein.}.
We study  what has been called the minimal supersymmetric Left-Right 
symmetric model (MSLRM) \cite{Aulakh:1998nn} 
with the gauge group $SU(3)_c$ $\otimes ~SU(2)_L$ $\otimes ~SU(2)_R$
$\otimes ~U(1)_{B-L}$ augmented by parity $P$ exchanging $L$ and $R$ 
sectors. 
Lee et al. \cite{Shafi:2007xp} have studied a similar model with the gauge
group $SU(4)_c$ $\otimes ~SU(2)_L$ $\otimes ~SU(2)_R$ and connected it
to cosmological phenomena, specifically inflation. Our discussion
differs in being specifically PeV scale.

In the MSLRM class of Left-Right symmetric models, spontaneous 
gauge symmetry breaking required to recover SM phenomenology also 
leads to observed parity breaking. However, for cosmological reasons 
it is not sufficient to ensure local breakdown of parity. We have 
earlier proposed \cite{Yajnik:2006kc} that the occurrence of the SM like 
sector globally is connected to the SUSY breaking effects from the 
hidden sector. Another approach to implementing the global uniformity of
parity breaking is to have terms induced by gauge symmetry breaking 
which signal explicit parity breaking \cite{Chang:1984uy, Sarkar:2007ic}.
This model has been dubbed MSLR{\rlap/P}. In earlier papers we have explored 
the overall cosmological setting for these models and traced issues such
as removal of unwanted relics and a successful completion of the first
order phase transition. Here we show that sufficient conditions exist
in the model to provide for the leptogenesis required to explain the
baryon asymmetry of the Universe.

A possible implementation of this idea follows the thermal leptogenesis
\cite{Fukugita:1986hr} route. This however has been shown to 
generically require 
the scale of  majorana neutrino mass, equivalently, in our model the 
scale of $B-L$ breaking to be $10^{11}$-$10^{13}$ GeV 
\cite{Davidson:2002qv, Hambye:2001eu}, 
with a more optimistic constraint $M_{B-L} >10^9$GeV 
\cite{Buchmuller:2003gz, Buchmuller:2004tu}. 
This situation is not improved 
\cite{Sahu:2004ir, Sahu:2005vu, Jeannerot:2005ah, Gu:2005gu} 
by assistance from cosmic string induced violation 
\cite{Stern:1985bg, Kawasaki:1987vg, Davis:1992fm} 
of lepton number \cite{Jeannerot:1996yi}. 
On the other hand, it has been shown \cite{Fischler:1990gn, Sahu:2004sb}
that the only real requirement imposed by Leptogenesis is that the presence
of heavy neutrinos should not erase lepton asymmetry generated by a given 
mechanism, possibly non-thermal. This places the modest bound $M_1> 10^4$GeV,
on the mass of the lightest  of the  heavy majorana neutrinos. A scenario which exploits this window and relies on supersymmetry is the ``soft leptogenesis", 
\cite{Grossman:2003jv, D'Ambrosio:2003wy, Boubekeur:2004ez, Chun:2005ms}
relying on the decay of scalar superpartners of neutrino and a high degree
of degeneracy \cite{Pilaftsis:1997jf} 
in the mass eigenvalues due to soft SUSY  breaking terms. 

Another possibility for leptogenesis arises from the fact that generically
the Left-Right breaking phase transition is intrinsically a first order phase 
transition. Due to the presence of lepton number violating processes,
the problem of leptogenesis then becomes analogous to that explored for 
the electroweak phase transition \cite{Cohen:1993nk},
provided a source for $CP$ asymmetry can be found. It has been shown 
\cite{Cline:2002ia}
that the domain
walls arising during the phase transition generically give spatially
varying complex masses to neutrinos. Here we explore the parameter space
required in the two variants of Left-Right symmetric model to ensure the 
required leptogenesis.

The paper is organized as follows. In the next sections \ref{sec:mslrm} and
\ref{sec:mslrps} we review the models being considered. In sec. \ref{sec:scbrk} 
we discuss the cosmological evolution characteristic of each of the models, 
along with the constraints that can be obtained on the soft parameters 
of the models by the demand that the phase transition is completed 
successfully. In \ref{sec:susylgen} we identify the soft parameters in the
model that can be constrained by the demand for soft leptogenesis. In sec.s 
\ref{sec:dwlgen} and \ref{sec:wallprofiles} we detail the mechanism of
leptogenesis by the domain wall (DW) structure of the phase transition and then
obtain numerical solutions which support the possibility of this mechanism
to operate in the two models.  Conclusions are summarized in sec.
\ref{sec:conclusion}.

\section{MSLRM} 
\label{sec:mslrm}
The standard Left-Right symmetric model is based on the gauge group
$SU(3)_c$ $\otimes SU(2)_L$ $\otimes SU(2)_R$ $\otimes U(1)_{B-L}$. 
The right handed charged leptons which were singlet in standard 
model (SM), form doublets with respective right handed neutrino species $\nu_R$ 
under $SU(2)_R$ in this model. 
In the same manner, the right handed up and down quarks of
each generation which were singlets in SM, form doublets under $SU(2)_R$.
The Higgs sector has two triplets ($\Delta$'s), and a bidoublet ($\Phi$). 
In minimal supersymmetric Left-Right model (MSLRM) \cite{Aulakh:1998nn},
the bidoublet is doubled to have non-vanishing Cabibo-Kobayashi-Maskawa matrix 
and the triplets are doubled for reasons of anomaly cancellation. The quark 
and leptonic sectors along with their quantum numbers are represented below.
\begin{eqnarray}
Q = (3,2,1,1/3), & \quad & Q_c = (3^*,1,2,-1/3), \nonumber  \\
L = (1,2,1,-1),  & \quad & L_c = (1,1,2,1),
\end{eqnarray}
where we have suppressed the generation index. The minimal set of Higgs 
superfields required is,
\begin{eqnarray} 
\Phi_i = (1,2,2,0),    & \quad & i = 1,2, \nonumber \\
\Delta = (1,3,1,2),    & \quad & \bar{\Delta} = (1,3,1,-2), \nonumber \\
\Delta_c = (1,1,3,-2), & \quad & \bar{\Delta}_c = (1,1,3,2).
\end{eqnarray}
Under discrete parity symmetry the fields are prescribed to transform as,
\begin{eqnarray}
Q \leftrightarrow Q_c^*, \quad & 
L \leftrightarrow L_c^*, \quad & 
\Phi_i \leftrightarrow \Phi_i^\dagger,  \nonumber \\
\Delta \leftrightarrow \Delta_c^*,  \quad & 
\bar{\Delta} \leftrightarrow \bar{\Delta}_c^*.
\label{eq:parity} 
\end{eqnarray}
However, this minimal model is unable to break parity spontaneously 
\cite{Kuchimanchi:1993jg, Kuchimanchi:1995vk}. 
A parity odd singlet solves this problem \cite{Cvetic:1985zp}, 
but this also breaks electromagnetic charge invariance 
\cite{Kuchimanchi:1993jg}. 
Breaking $R$ parity and introducing non-renormalizable terms solves 
this problem. A more appealing 
way out is to introduce a pair of scalar triplets ($\Omega, \Omega_c$),
which are even under parity viz., $\Omega \leftrightarrow \Omega_c^*$
\cite{Aulakh:1997ba, Yajnik:2006kc, Sarkar:2007ic}. 
The quantum numbers for the two fields are,
\begin{equation} 
\Omega = (1,3,1,0), \qquad \Omega_c = (1,1,3,0) ~. 
\end{equation}
The superpotential for this model was given in 
\cite{Aulakh:1997ba}. 
It is almost the same as the superpotential given later in this paper, in 
sec. \ref{sec:mslrps}, eq. (\ref{eq:imsuperpot}) from which it can 
be obtained with $\Omega_c$  replaced by $-\Omega_c$. 
Since in this class of models, we consider supersymmetry to be broken only 
at the electroweak scale, we can safely employ the $F$-flatness and 
$D$-flatness conditions to obtain the vacua of the theory. The $F$ and $D$
flat conditions for MSLRM are given in ref \cite{Aulakh:1997ba} 
and again are similar in nature to the one we have worked out in 
appendix \ref{sec:flatness} for the modified version of this model 
discussed in sec. \ref{sec:mslrps}. These $F$ and $D$ flat conditions 
imply the existence of the following set of vacuum expectation values 
(vev's) for the Higgs fields as one of the possibilities.
\begin{equation} 
\begin{array}{ccc}
\langle \Omega \rangle = 0, & \qquad
\langle \Delta \rangle = 0, & \qquad
\langle \bar{\Delta} \rangle = 0, \\ [0.25cm]
\langle \Omega_c \rangle = 
\begin{pmatrix}
\omega_c & 0 \\
0 & - \omega_c
\end{pmatrix}, & \qquad
\langle \Delta_c \rangle =
\begin{pmatrix}
0 & 0 \\
d_c & 0
\end{pmatrix}, & \qquad
\langle \bar{\Delta}_c \rangle =
\begin{pmatrix}
0 & \bar{d}_c \\
0 & 0
\end{pmatrix}.
\label{eq:rhvev} 
\end{array}
\end{equation}
The stages of breaking required to implement parity breaking and avoid
electromagnetic charge breaking vacua, are as follows: 
first the $\Omega$'s get a vev at a scale $M_R$, which breaks $SU(2)_R$ to
its subgroup $U(1)_R$, but conserving $B-L$ charge. 
At a lower scale $M_{B-L}$, the triplets get vev
to break $U(1)_R$ $\otimes U(1)_{B-L}$
to $U(1)_Y$. Thus, at low scale MSLRM
breaks exactly to minimal supersymmetric standard model (MSSM).

From the $F$ and $D$ 
flatness conditions we are led to the following solution for the vev's 
\cite{Aulakh:1997ba, Yajnik:2006kc, Sarkar:2007ic}
%
\begin{eqnarray} 
\left|\omega\right| &=& \left|\frac{m_\Delta}{a} \right|
\equiv M_R , \nonumber \\
|d|=|\bar{d}| &=&
\left| \frac{2m_\Delta m_\Omega}{a^2}\right|^{1/2} \equiv M_{B-L}
\label{eq:vevvalues}
\end{eqnarray}
For parity breakdown we must have $M_R \gg M_{B-L}$, which is accomplished
if we have $m_\Delta \gg m_\Omega$. 
If the mass scale $m_\Omega$ originates from the soft terms, then we can
accept the approach of Ref \cite{Aulakh:1997fq} 
that $m_\Omega \simeq M_{EW}$. This
in turn would mean that $m_\Omega$ is of the same order as the gravitino
mass $m_{3/2}$. This leads us to the relation
\begin{equation}
M_{B-L}^2 \simeq M_R M_{EW}. 
\label{eq:scale-rel} 
\end{equation}
Thus, we have only one effective new mass scale, either $M_R$ or $M_{B-L}$.
Now if we consider $M_{B-L} \sim 10^4$ GeV, then $M_R \sim 10^6$ GeV.
On the other hand, $M_{B-L} \sim 10^6$ GeV, if we choose $M_R$ to have the 
largest possible value $ \sim \sqrt{M_{Pl}M_{EW}} \sim 10^{10}$ GeV, beyond 
which non-renormalizable terms will relevant. Thus the model is workable in 
a wide range of values, but the lower range values make the model verifiable
in the colliders.

The above solution for the vev's, is not unique. Due to Left-Right symmetric 
nature of the original theory, an alternative set of vev's permitted by the 
$F$ and $D$ flatness conditions are,
\begin{equation} 
\begin{array}{ccc}
\langle \Omega \rangle = 
\begin{pmatrix}
\omega & 0 \\
0 & - \omega
\end{pmatrix}, & \qquad
\langle \Delta \rangle =
\begin{pmatrix}
0 & 0 \\
d & 0
\end{pmatrix}, & \qquad
\langle \bar{\Delta} \rangle =
\begin{pmatrix}
0 & \bar{d} \\
0 & 0
\end{pmatrix}, \\[0.50cm]
\langle \Omega_c \rangle = 0, & \qquad
\langle \Delta_c \rangle = 0, & \qquad
\langle \bar{\Delta}_c \rangle = 0.
\label{eq:lhvev} 
\end{array}
\end{equation}
Due to the possibility of alternative set of Higgs vacua, in the early universe,
parity breakdown does not select unique ground state and  formation
of domain walls (DW) is inevitable. As this contradicts present observable
cosmology the model must have an inbuilt asymmetry to remove the domain 
walls. Since the superpotential doesn't allow such asymmetry in the
present model, we depend on the soft terms to do the job. 

The mechanism which induces the soft terms can arise 
due to gravitational effects in the gravity mediated supersymmetry breaking. 
In gauge mediated supersymmetry breaking (GMSB), the soft terms can arise 
due to the messenger sector, the hidden sector or both.
In the next section \ref{sec:mslrps} however, we look for an alternative 
possibility for the breaking parity, which arises naturally out of the 
Higgs sector.

\section{MSLR\rlap/P} 
\label{sec:mslrps} 

In this section we consider another possibility for parity 
breaking which takes place within the Higgs sector. The idea was first
considered by Chang {\it et al.} \cite{Chang:1984uy},
for the non-susy model 
$SU(3)_c$ $\otimes SU(2)_L$ $\otimes SU(2)_R$ $\otimes U(1)_{B-L}$ 
$\otimes P$ where $P$ denotes parity symmetry. To break parity an extra 
Higgs singlet $\eta$ which is odd under $P$ parity was introduced .i.e 
$\eta \leftrightarrow - \eta$. 
As such the potential of the model has a term of the form
\begin{equation}
V_{\eta\Delta} \sim M \eta (\Delta_L^\dagger \Delta_L 
- \Delta_R^\dagger \Delta_R),
\end{equation} 
where the notation is self-evident. 
Thus, when at a high scale $M_P$, the singlet $\eta$ gets a vev, the 
effective masses of the left and right triplet Higgs masses become different,
thus explicitly breaking $P$ parity, without affecting $SU(2)_R$. 
However, in SUSY, a parity odd singlet in the theory would generate the 
problems of charge breaking vacua as discussed by Kuchimanchi and Mohapatra
\cite{Kuchimanchi:1993jg}.
To avoid this, but to implement the idea of Chang {\it et al.}
we propose an alternative SUSY model based on the group $SU(3)_c$ 
$\otimes SU(2)_L$ $\otimes SU(2)_R$ $\otimes U(1)_{B-L}$ $\otimes P$ with 
a pair triplets $(\Omega, \Omega_c)$ which are odd under parity. 
This model was discussed in an earlier paper \cite{Sarkar:2007ic}
and was named MSLR\rlap/P . Under parity,
\begin{eqnarray}
Q \leftrightarrow Q_c^*, \quad & 
L \leftrightarrow L_c^*, \quad & 
\Phi_i \leftrightarrow \Phi_i^\dagger,  \nonumber \\
\Delta \leftrightarrow \Delta_c^*,  \quad & 
\bar{\Delta} \leftrightarrow \bar{\Delta}_c^*, \quad & 
\Omega \leftrightarrow -\Omega_c^*.
\label{eq:imparity}
\end{eqnarray}
The superpotential for this parity symmetry becomes,
\begin{eqnarray}
 W_{LR}&=& {\bf h}_l^{(i)} L^T \tau_2 \Phi_i \tau_2 L_c 
 + {\bf h}_q^{(i)} Q^T \tau_2 \Phi_i \tau_2 Q_c 
 +i {\bf f} L^T \tau_2 \Delta L 
 +i {\bf f} L^{cT}\tau_2 \Delta_c L_c \nonumber \\
&& + ~m_\Delta  {\rm  Tr}\, \Delta \bar{\Delta} 
  + m_\Delta  {\rm Tr}\,\Delta_c \bar{\Delta}_c
  + \frac{m_\Omega}{2} {\rm Tr}\,\Omega^2 
  + \frac{m_\Omega}{2} {\rm Tr}\,\Omega_c^2 \nonumber \\
&& + ~\mu_{ij} {\rm Tr}\,  \tau_2 \Phi^T_i \tau_2 \Phi_j 
  +a {\rm Tr}\,\Delta \Omega \bar{\Delta}
  -a {\rm Tr}\,\Delta_c \Omega_c \bar{\Delta}_c \nonumber \\
& &  + ~\alpha_{ij} {\rm Tr}\, \Omega  \Phi_i \tau_2 \Phi_j^T \tau_2 
  - \alpha_{ij} {\rm Tr}\, \Omega_c  \Phi^T_i \tau_2 \Phi_j \tau_2 ~,
\label{eq:imsuperpot}
\end{eqnarray} 
where color and flavor indices have been suppressed.
Further, ${\bf h}^{(i)}_{q}  =  {{\bf h}^{(i)}_{q}}^\dagger $,
${\bf h}^{(i)}_{l}  =  {{\bf h}^{(i)}_{l}}^\dagger $,  
$\mu_{ij}  =  \mu_{ji} = \mu_{ij}^*$,
$\alpha_{ij} = -\alpha_{ji}$. Finally, ${\bf f}$, ${\bf h}$
are real symmetric matrices with respect to flavor indices. 

The $F$ and $D$ flatness conditions derived from this superpotential are 
presented in appendix \ref{sec:flatness}. However, the effective
potential for the scalar fields which is determined from modulus
square of the $D$ terms remains the same as for the MSLRM at least 
for the form of the ansatz of the vev's we have chosen. 
As such the resulting solution for the vev's remains identical 
to eq. (\ref{eq:vevvalues}). The difference in the effective
potential shows up in the soft terms as will be shown later. 
Due to soft terms, below  the scale $M_R$ the effective 
mass contributions to $\Delta$ and $\bar{\Delta}$ become
larger than those of $\Delta_c$ and $\bar{\Delta}_c$. The cosmological
consequence of this is manifested after the $M_{B-L}$ phase transition
when the $\Delta$'s become massive. Unlike MSLRM where the DW 
are destabilized only after the soft terms become significant, i.e., 
at the electroweak scale, the DW in this case become unstable immediately
after $M_{B-L}$. Leptogenesis therefore commences immediately below 
this scale and the scenario becomes qualitatively different from 
that for the MSLRM.

In the next section we elaborate in detail the areas where the two models 
MSLRM and MSLR\rlap/P differ from the cosmological point of view.

\section{Cosmology of breaking}
\label{sec:scbrk} 
In this section we recapitulate the cosmology of these models.
In the two models MSLRM and MSLR\rlap/P the stages of breaking  
are slightly different as shown in Table (\ref{tab:pattern}). 
Domain walls form in both the models at the scale
$M_R$, when the $\Omega$ fields get vev. These DW come to dominate the 
evolution of the Universe and is responsible for the onset of a secondary
inflation. This secondary inflation removes gravitinos and other relic
abundances which were regenerated during the reheating stage after the 
primordial inflation ended \cite{Yajnik:2006kc, Sarkar:2007ic}.
%
\begin{table}[!htp]
\setlength{\extrarowheight}{1.5mm} 
\begin{tabular}{c|c|c|c|c}
\hline
Cosmology & Scale & Symmetry Group & MSLR\rlap/P & MSLRM \\
&&&(GeV)&(GeV) \\ [1.5mm] \hline \hline
\multirow{3}{1.75in}{%
$\Omega$ or $\Omega_c$ get vev. \\
Onset of wall dominated \\ secondary inflation.} & &
$SU(3)_c \otimes SU(2)_L \otimes SU(2)_R \otimes U(1)_{B-L}$ & &
\\
& $M_R$ & $\downarrow$ & $10^6$ & $10^6$  \\
&&&& 
\\ [1.5mm] \hline
\multirow{2}{1.75in}{%
Higgs triplet $(\Delta's)$ \\ get vev } & &
$SU(3)_c \otimes SU(2)_L \otimes U(1)_R \otimes U(1)_{B-L}$ && \\
& $M_{B-L}$ & $\downarrow$ & $10^4$ & $10^4$ \\ [1.5mm] \hline
\multirow{2}{1.75in}{%
End of inflation and \\ beginning of L-genesis} &
$M_{B-L}$
&& $10^4$ & --- \\ \cline{2-5} 
& $M_S$ & & --- & $10^3$ \\ [1.5mm] \hline
\multirow{2}{*}{SUSY breaking} & & 
$SU(3)_c \otimes SU(2)_L \otimes U(1)_Y$ (SUSY) & & \\ 
& $M_S$ & $\downarrow$ & $10^3$ & $10^3$ \\ [1.5mm] \hline
\multirow{2}{1.75in}{Wall disappearance \\ temperature} & 
\multirow{2}{*}{$T_D$} & & 
\multirow{2}{*}{$10 - 10^3$} & 
\multirow{2}{*}{$10 - 10^2$} \\
&&&& \\ [1.5mm] \hline
\multirow{2}{1.75in}{Secondary reheat \\ temperature} & 
\multirow{2}{*}{$T^s_R$} & & 
\multirow{2}{*}{$10^3 - 10^4$} & 
\multirow{2}{*}{$10^3$} \\
&&&&\\[1.5mm] \hline
\multirow{2}{*}{Electroweak breaking} & &
$SU(3)_c \otimes SU(2)_L \otimes U(1)_Y$ (non-SUSY)& & 
\\ & $M_{EW}$ & $\downarrow$ & $10^2$ & $10^2$ \\[1.5mm] \hline
Standard Model & & $SU(3)_c \otimes U(1)_{EW}$ & & 
\\ [1.5mm] \hline \hline
\end{tabular}
\caption{Pattern of symmetry breaking and the slightly different sequence
of associated cosmological events in the two classes of models}
\label{tab:pattern}
\end{table}
At the scale 
$M_{B-L}$, the triplet $\Delta$'s get vev. At this epoch the effective mass 
of the left-handed $\Delta$'s is essentially different than those of 
right-handed $\Delta$'s in MSLR\rlap/P. As such at this stage DW are 
destabilized and leptogenesis begins in MSLR\rlap/P unlike in MSLRM.
SUSY breaking is mediated from the hidden sector to the visible sector in 
both the models at the scale $M_S$. The soft terms which become 
relevant at this scale break the parity in MSLRM. Thus the DW become 
destabilized in MSLRM at $M_S$, thus beginning the process of leptogenesis.
The walls finally disappear in MSLR\rlap/P at a scale $T_D \sim 10-10^3$ GeV
and in MSLRM at $T_D \sim 10-10^2$ GeV. Subsequently standard cosmology 
takes over after this.

A handle on the explicit symmetry breaking parameters 
of the two models can be obtained by noting that there should exist
sufficient wall tension for the walls to disappear before
a desirable temperature scale $T_D$. It has been observed in
\cite{Preskill:1991kd} 
that the free energy density difference $\delta \rho$
between the vacua, which determines the pressure difference across 
a domain wall should be of the order 
\begin{equation}
\delta \rho \sim T_D^4
\label{eq:dr_Td_rel}
\end{equation}
in order for the DW structure to disappear at the scale $T_D$. 

\subsection{Consistent cosmology : MSLRM}
\label{sec:remove-mslrm}
The possible source for breaking the parity symmetry of the
MSLRM lies in soft terms with the assumption that the 
hidden sector, or in case of GMSB also
perhaps the messenger sector does not obey the parity of the
visible sector model. For gravity mediated breaking this can
be achieved in a natural way since a discrete symmetry can be 
generically broken by gravity effects. We present the possible
soft terms for MSLRM below.
\begin{eqnarray} 
\mathcal{L}_{soft} &=& 
\alpha_1 \textrm{Tr} (\Delta \Omega \Delta^{\dagger}) +
\alpha_2 \textrm{Tr} (\bar{\Delta} \Omega \bar{\Delta}^{\dagger}) +
\alpha_3 \textrm{Tr} (\Delta_c \Omega_c \Delta^{\dagger}_c) + 
\alpha_4 \textrm{Tr} (\bar{\Delta}_c \Omega_c \bar{\Delta}^{\dagger}_c) ~~~~~
\nonumber \\ 
&& + ~m_1^2 \textrm{Tr} (\Delta \Delta^{\dagger}) +
m_2^2 \textrm{Tr} (\bar{\Delta} \bar{\Delta}^{\dagger}) + 
m_3^2 \textrm{Tr} (\Delta_c \Delta^{\dagger}_c) +
m_4^2 \textrm{Tr} (\bar{\Delta}_c \bar{\Delta}^{\dagger}_c) 
\nonumber \\
&& + ~\beta_1 \textrm{Tr} (\Omega \Omega^{\dagger}) +
\beta_2 \textrm{Tr} (\Omega_c \Omega^{\dagger}_c) ~.
\label{eq:omega} 
\end{eqnarray} 

We can determine the differences between the relevant soft parameters
for a range of permissible values of $T_D$. 

\begin{table}
\begin{center} 
{\setlength\extrarowheight{1.5mm}
\begin{tabular}{cc|c|c|c|c|c} 
\hline
$T_D/$GeV & $\sim$ & $10^{-1}$ & $1$ & $10$ 
& $10^2$ & $10^3$ \\ \hline \hline
$(m^2 - m^{2\prime})/\mbox{GeV}^2$ & $\sim$ 
& $10^{-12} $  & $10^{-8}  $ & $10^{-4}  $  & $1$  & $10^{4} $\\ [1mm]
$(\beta_1 - \beta_2)/\mbox{GeV}^2 $ & $\sim$ 
& $10^{-16} $  & $10^{-12} $  & $10^{-8}  $ & $10^{-4} $ & 
$1$ \\ [1mm] \hline \hline
\end{tabular} }
\end{center}
\caption{Differences in values of soft supersymmetry breaking parameters of 
MSLRM, for a range of domain wall decay temperature values $T_D$. The 
differences signify the extent of parity breaking. 
}
\label{tab:DWalls}
\end{table}

In Table \ref{tab:DWalls} we have taken $d \sim 10^4 $ GeV, 
$\omega \sim 10^6 $ GeV and $T_D$ in the range 
$100 \textrm{ MeV} - 10 \textrm{ GeV}$ \cite{Kawasaki:2004rx}.
The above differences between the values in the left and right sectors is a 
lower bound on the soft parameters and is very small. Larger values would
be acceptable to low energy phenomenology. However if we wish to retain the
connection to the hidden sector, and have the advantage of secondary 
inflation we would want the differences to be close to this bound.
As pointed out in \cite{Preskill:1991kd, Dine:1993yw}
an asymmetry $\sim 10^{-12}$ 
is sufficient to lift the degeneracy between the two sectors.

\subsection{Consistent cosmology : MSLR\rlap/P}
\label{sec:remove-pslash}
In this model parity breaking is achieved spontaneously within
the observable sector below the scale $M_R$ at which the $\Omega$
fields acquire vev's. However the breaking is not manifested in the
vacuum till the scale $M_{B-L}$ where the $\Delta$ fields acquire 
vev's. For simplicity 
we assume that the hidden sector  responsible for SUSY breaking
does not contribute parity breaking terms. This is reasonable
since even if the hidden sector breaks this parity the corresponding 
effects are suppressed by the higher scale of breaking
and in the visible sector the parity breaking effects are
dominated by the explicit mechanism proposed. Thus at a scale
above $M_R$ but at which SUSY is broken 
in the hidden sector we get induced soft terms respecting this
parity. Accordingly, for the Higgs sector the parameters can be 
chosen such that 
\begin{eqnarray} 
\mathcal{L}_{soft} &=& 
\alpha_1 \textrm{Tr} (\Delta \Omega \Delta^{\dagger}) -
\alpha_2 \textrm{Tr} (\bar{\Delta} \Omega \bar{\Delta}^{\dagger}) -
\alpha_1 \textrm{Tr} (\Delta_c \Omega_c \Delta^{\dagger}_c) +
\alpha_2 \textrm{Tr} (\bar{\Delta}_c \Omega_c \bar{\Delta}^{\dagger}_c) ~~~~~
\label{eq:imsigNdel} 
\nonumber \\ 
&& + ~m_1^2 \textrm{Tr} (\Delta \Delta^{\dagger}) +
m_2^2 \textrm{Tr} (\bar{\Delta} \bar{\Delta}^{\dagger}) + 
m_1^2 \textrm{Tr} (\Delta_c \Delta^{\dagger}_c) +
m_2^2 \textrm{Tr} (\bar{\Delta}_c \bar{\Delta}^{\dagger}_c) 
\label{eq:imdelta} 
\nonumber \\
&& + ~\beta \textrm{Tr} (\Omega \Omega^{\dagger}) +
\beta \textrm{Tr} (\Omega_c \Omega^{\dagger}_c) ~.
\label{eq:imomega} 
\end{eqnarray} 
These terms remain unimportant at first due to the key assumption 
leading to MSSM as the effective low energy theory. The SUSY
breaking effects become significant only at the electroweak scale.
However, below the scale $M_R$, $\Omega$ and $\Omega_c$ acquire vev's
given by eq. (\ref{eq:rhvev}) or (\ref{eq:lhvev}).
Further, below the scale $M_{B-L}$ the $\Delta$ fields acquire vev's
and become massive. The combined contribution from the superpotential 
and the soft terms to the $\Delta$ masses now explicitly encodes the
parity breaking,
\begin{equation} 
\begin{array}{cc}
\mu^2_\Delta = M^2_\Delta + \alpha_1 \omega, & \qquad 
\mu^2_{\Delta_c} = M^2_\Delta - \alpha_1 \omega, \\ [3mm]
\mu^2_{\bar{\Delta}} = M^2_\Delta + \alpha_2 \omega,  & \qquad 
\mu^2_{\bar{\Delta}_c} = M^2_\Delta - \alpha_2 \omega. 
\end{array}
\label{eq:thiggs_mc} 
\end{equation} 
where $M^2_\Delta$ is the common contribution from the superpotential.
The difference in free energy across the domain wall is now 
dominated by the differential contribution to the $\Delta$ masses
\begin{equation} 
\delta\rho_\alpha \equiv 2(\alpha_1 + \alpha_2) \omega d^2, 
\label{eq:lrasymm} 
\end{equation} 
where we have considered 
$\omega_c \sim \omega$, $d \sim \bar{d} \sim d_c \sim \bar{d}_c$.
Now using  eq (\ref{eq:dr_Td_rel}) for a range of temperatures 
$(T_D \sim 10^2 \text{ GeV} - 10^4 \text{ GeV})$,
determines the corresponding range of values
of coupling constants as 
\begin{equation} 
(\alpha_1 + \alpha_2) \sim 10^{-6} - 10^{2} \text{ GeV},
\label{eq:a_diff} 
\end{equation} 
where we have considered $|\omega| \simeq M_R$, $|d| \simeq M_{B-L}$.

\section{Supersymmetry and leptogenesis}
\label{sec:susylgen}
The supersymmetric Left-Right symmetric models considered here do not favor 
generic thermal leptogenesis from decay of heavy majorana neutrinos for an 
intriguing  reason. $B-L$ asymmetry  
in  the form of fermion chemical potential 
is guaranteed to remain zero in the model until the gauged $B-L$ 
symmetry breaks spontaneously. As can be seen, a generic consequence of
symmetry breaking in both the models is a relation among the various mass 
scales $M^2_{B-L} \simeq M_{EW} M_R$. Thermal Leptogenesis
requires $M_{B-L}$ to be larger than $10^{11}$-$10^{13}$ GeV, which pushes 
$M_R$ into the Planck scale in light of the above formula. A more 
optimistic constraint $M_{B-L} >10^9$GeV 
\cite{Buchmuller:2003gz, Buchmuller:2004tu} 
requires Left-Right symmetry to be essentially Grand Unified theory.

However, supersymmetry provides new channels for thermal leptogenesis
via out of equilibrium decay of scalar superpartners of leptons 
\cite{D'Ambrosio:2003wy, Grossman:2003jv, Boubekeur:2004ez}.
Leptogenesis from scalar sector is free of strong constraints
on the Yukawa couplings as happens in thermal leptogenesis from fermion
decay \cite{Hambye:2001eu}. 
In the mechanism to be discussed, the sneutrino splits 
into two distinct mass eigenstates due to soft supersymmetry breaking terms. 
The relevant terms in the superpotential are
\begin{equation}
 W_{leptonic}= {\bf h}_l^{(i)} L^T \tau_2 \Phi_i \tau_2 L_c 
 +i {\bf f} L^T \tau_2 \Delta L 
 +i {\bf f} L^{cT}\tau_2 \Delta_c L_c 
\end{equation}
The relevant soft terms $(V_{ls})$ in our model are given by
\begin{equation}
V_{ls} = A h^{(i)} \widetilde{L}^T \tau_2 \Phi_i \tau_2 
\widetilde{L}_c 
+ i B f \widetilde{L}^T \tau_2 \Delta \widetilde{L}
+ i B^\prime f \widetilde{L}_c^{T} \tau_2 \Delta_c \widetilde{L}_c
+ \widetilde{m}^2 \widetilde{L}^\dagger \widetilde{L}
+ \widetilde{m}^2 \widetilde{L}_c^\dagger \widetilde{L}_c
\label{eq:vls}
\end{equation} 
Mixing between the two states of sneutrino generates the CP violation.
 
Consider the generic model introduced by \cite{D'Ambrosio:2003wy},
where the superpotential is given by 
\begin{equation}
W = h L H N + \frac{1}{2}MNN ,
\end{equation} 
where, $L$, $H$ and $N$ are the left-handed lepton doublet, the Higgs and the 
right handed neutrino respectively. Here we have omitted the generation index 
for simplicity of notation. The SUSY soft breaking terms are given by,
\begin{equation}
V_\textrm{soft} = \left[A h \widetilde{L} H \widetilde{N} 
+ \frac{1}{2} B M \widetilde{N} \widetilde{N} + \textrm{h.c} \right] 
+ \widetilde{m}^2 \widetilde{N}^\dagger \widetilde{N}
\end{equation} 
%
The mixing between the two eigenstates in the decay of the right-handed 
sneutrino $(\widetilde{N})$ produces the required CP violation $(\epsilon)$. 
The two eigenstates $\widetilde{N}_1$ and $\widetilde{N}_2$ of the sneutrino,
$\widetilde{N} = (\widetilde{N}_1 + i\widetilde{N}_2)/\sqrt{2} $
are given by
\begin{equation}
M^2_{\widetilde{N}_{1,2}} = M^2 + \widetilde{m}^2 \pm B M
\end{equation} 
Due to the near degeneracy of these masses
the CP asymmetry can be large.
The mechanism has been studied in detail in \cite{Chun:2005tg} 
where it is shown that the constraint on the soft parameter $B$ is
\begin{equation}
B \sim \Gamma \sim 0.1 \textrm{ eV} 
\left( \frac{m_\nu}{0.05 \textrm{ eV}} \right) 
\left( \frac{M}{\textrm{TeV}} \right)
\end{equation} 
This is the same as the $B$ parameter in our model introduced in 
eq. (\ref{eq:vls}). In \cite{Chun:2005tg}
it is shown that this constraint 
can be corroborated by collider experiments involving $Z'$ decays.
The $Z'$ sector of the model we are considering is similar and
similar collider constraints are applicable. 

Further, we see that the $B$ required is $O(10^{-12})$ relative to 
the electroweak scale. This smallness of the value is possible
in certain scenarios \cite{chun:2004eq} 
and is expected in
models of hidden sector supersymmetry breaking. Here we see a 
correspondence between the smallness of this parameter and the
parameters in the Higgs sector as determined from the cosmological 
constraint of disappearance of the DW summarized in sec. \ref{sec:remove-mslrm}.
This is a strong indication that we may be able to test the validity
of MSLRM by ascertaining its hidden sector breaking scheme and
correlating the two cosmological requirements determined from 
smallness of otherwise unrelated parameters arising from the same mechanism.

\section{Leptogenesis from first order phase transition}
\label{sec:dwlgen}
In addition to the resonant leptogenesis considered in previous section,
the models considered here also include natural possibility of 
non-thermal leptogenesis.
The spontaneous breaking of a discrete symmetry automatically makes the
Left-Right symmetry breaking phase transition a first order phase transition.
The idea is similar to electroweak baryogenesis proposals 
\cite{Nelson:1991ab, Cohen:1993nk}
where there are spontaneously formed bubbles which expand to complete
the phase transition, a mechanism also considered in the case of 
Left-Right symmetric model in \cite{Frere:1992bd}. 
The dynamics of Left-Right breaking phase transition considered
here takes into  account that due to parity symmetry of the
theory both Right-like (unbroken $SU(2)_R$), and Left-like 
(unbroken $SU(2)_L$) domains are liable to occur
at the phase transition. In the models considered here
parity is unbroken at the first stage of the symmetry breaking.
The phase transition is accompanied by the spontaneous formation 
of domain walls separating Left-like and Right-like regions.
At a lower scale when parity breaking is signalled, the walls 
sweep through the Universe ensuring global choice  of a unique 
phase everywhere. The domain walls move irreversibly during this 
epoch, 
thereby eliminating the energetically unfavorable phase and 
providing time irreversibility.

Consider the interaction of neutrinos with the L-R wall, which is
encroaching on the energetically disfavored phase.  The left-handed
neutrinos, $\nu_\Ls$,  are massive in this domain, whereas they are
massless in the phase behind the wall. More precisely, as per see-saw
mechanism, $\nu_\Ls$ constitute the principal component of the 
heavy mass eigenstate in front of the wall but become principal 
component of the light eigenstate behind the
wall, and it is the $\nu_\Ls$ whose fate we keep track of. 
To get leptogenesis, one needs an asymmetry in the reflection and
transmission coefficients from the wall between $\nu_\Ls$ and its CP
conjugate $(\nu_\Ls^c)$.  This can happen if a CP-violating condensate
exists in the wall. This comes from the Dirac mass terms 
as discussed in \cite{Joyce:1994fu, Joyce:1994zn, Cline:1997vk,
Cline:2000nw, Cline:2000kb}.
Then there will be a preference
for transmission of, say, $\nu_\Ls$. The corresponding excess of
antineutrinos $(\nu_\Ls^c)$ reflected in front of the wall will quickly
equilibrate with $\nu_\Ls$ due to helicity-flipping scatterings, whose
amplitude is proportional to the large Majorana mass.  However the
transmitted excess of $\nu_\Ls$ survives because it is not coupled to
its CP conjugate in the region behind the wall, where the majorana
mass contribution from $\vev {\Delta}$ and $\vev {\Delta_c}$ vanishes.

A quantitative analysis of this effect can be made either in the
framework of quantum mechanical reflection, valid for domain walls
which are narrow compared to the particles' thermal de Broglie
wavelengths, or using the classical force method  
\cite{Joyce:1994fu, Joyce:1994zn, Cline:1997vk, Cline:2000nw, Cline:2000kb}
which gives the dominant contribution for walls
with larger widths. We adopt the latter here. The thickness
of the wall depends on the shape of the effective quartic potential
and we shall here treat the case of thick walls. Further, 
we assume that the potential energy difference between 
the two kinds of vacua is small, for example suppressed by Planck 
scale effects. In this case the pressure difference across the 
phase boundary is expected to be small, leading to slowly moving walls. 
The classical CP-violating force of 
the condensate on a fermion (in our case a neutrino) with
momentum component $p_x$ perpendicular to the wall can be shown to be
\be
\label{eq:force}
F = \pm {\rm sign}(p_x)\frac{1}{2E^2}\left(m_\nu^2(x) 
\chi'(x)\right)'.
\ee
The sign depends on whether the particle is $\nu_\Ls$ or $\nu_\Ls^c$, 
$m_\nu^2(x)$ is the position-dependent mass, \(E\) the 
energy  and $\chi$ is
the spatially varying CP-violating phase. One can then derive
a diffusion equation for the chemical potential $\mu_\Ls$ of the
$\nu_\Ls$ as seen in the wall rest frame:
\be
\label{eq:diffeq}
-D_\nu \mu_\Ls'' - v_w \mu_\Ls'
+ \theta(x)\, \Gamma_{\rm hf}\,\mu_\Ls = S(x).
\ee
Here $D_\nu$ is the neutrino diffusion coefficient,
$v_w$ is the velocity of the wall, taken to be moving in the $+x$
direction, $\Gamma_{\rm hf}$ is the rate of helicity
flipping interactions taking place in front of the wall (hence
the step function $\theta(x)$), and $S$ is the source term,
given by
\be
\label{eq:source}
   S(x) = - \frac{v_w D_\nu}{\vev{\vec v^{\,2}}} 
	\vev{v_x F(x)}',
\ee
where $\vec v$ is the neutrino velocity and
the angular brackets indicate thermal averages.
The net lepton number excess can then be calculated from
the chemical potential resulting as the solution of 
eq.\ (\ref{eq:diffeq}).

In order to use this formalism it is necessary to establish
the presence of a position-dependent phase $\chi$. This
is what we turn to in the following discussion of the nature 
of domain walls in the L-R model.

\section{Wall profiles and CP violating condensate}
\label{sec:wallprofiles}

In order for nontrivial effects to be mediated by the
walls, the fermion species of interest should get a 
space-dependent mass from the wall. Furthermore, the CP-violating
phase $\chi$ should also possess a nonvanishing gradient in the
wall interior. We study the minimization of the total
energy functional of the scalar sector with this in mind.

The vev's introduced in eq. (\ref{eq:rhvev}) are in general complex. Some of 
them can be rendered real by global $SU(2)$  transformations 
\cite{Deshpande:1990ip, Chen:2004ww} 
\begin{equation}
U_L = \begin{pmatrix} 
e^{i\gamma_L} & 0 \\
0 & e^{-i\gamma_L}
\end{pmatrix},
\qquad
U_R = \begin{pmatrix} 
e^{i\gamma_R} & 0 \\
0 & e^{-i\gamma_R}
\end{pmatrix}
\end{equation} 
according to
\begin{eqnarray}
\Phi_1 \rightarrow U_L \Phi_1 U_R^\dagger, & \qquad &
\Phi_2 \rightarrow U_L \Phi_2 U_R^\dagger, \\ 
\Delta \rightarrow U_L \Delta U_L^\dagger, & \qquad &
\bar{\Delta} \rightarrow U_L \bar{\Delta} U_L^\dagger, \\ 
\Delta_c \rightarrow U_R \Delta_c U_R^\dagger, & \qquad &
\bar{\Delta}_c \rightarrow U_R \bar{\Delta}_c U_R^\dagger, \\ 
\Omega \rightarrow U_L \Omega U_L^\dagger, &\qquad&
\Omega_c \rightarrow U_R \Omega_c U_R^\dagger.
\end{eqnarray} 
The vev's of the triplets $\Omega$ and $\Omega_c$ being diagonal are not 
affected by these transformations. Their phases if any do not enter fermion
or sfermion masses. We choose their phases to be real. This leaves us with 
16 degrees of freedom in the Higgs sector. These can be parameterized
by allowing three of the vev's in the four $\Delta$ fields and three of 
the vev's in the two bidoublets $\Phi$ to be complex. Here we present a
simpler model. As shown in eq.s (\ref{eq:simvev}) and (\ref{eq:simphi}), 
only two of the vev's are chosen to be complex, viz., the $\Delta$ and upper 
component of $\Phi_1$. The parameters $\alpha_{ij}$ reduce to a single
value $\alpha$ times the anti-symmetric matrix $\epsilon_{ij}$, and 
all the four values of $\mu_{ij}$ are chosen to be the same value $\mu$.
We have also studied
the model with all the allowable phases to be non-zero and find that it
does not result in any substantial improvement to the required condition 
for leptogenesis. The simpler model contains the minimal features to 
reproduce all the  essential features required for leptogenesis. 

\begin{figure}[!htp]
\begin{center}
\includegraphics[width=0.90\textwidth]{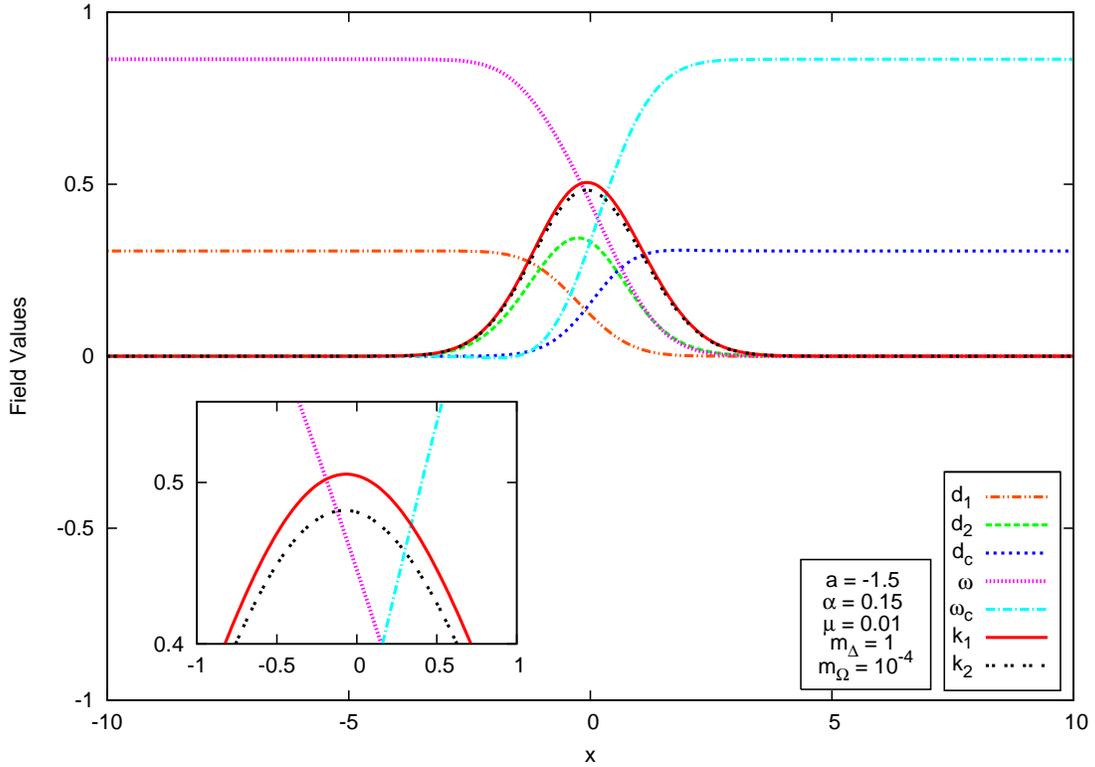}
\caption{Domain wall with $CP$ violating condensate in MSLRM. The
inset magnifies the behaviour of the $k_1$ and $k_2$ near their maximum value.}
\label{fig:panelMSLRM}
\end{center}
\end{figure}

\begin{figure}[!htp]
\begin{center}
\includegraphics[width=0.90\textwidth]{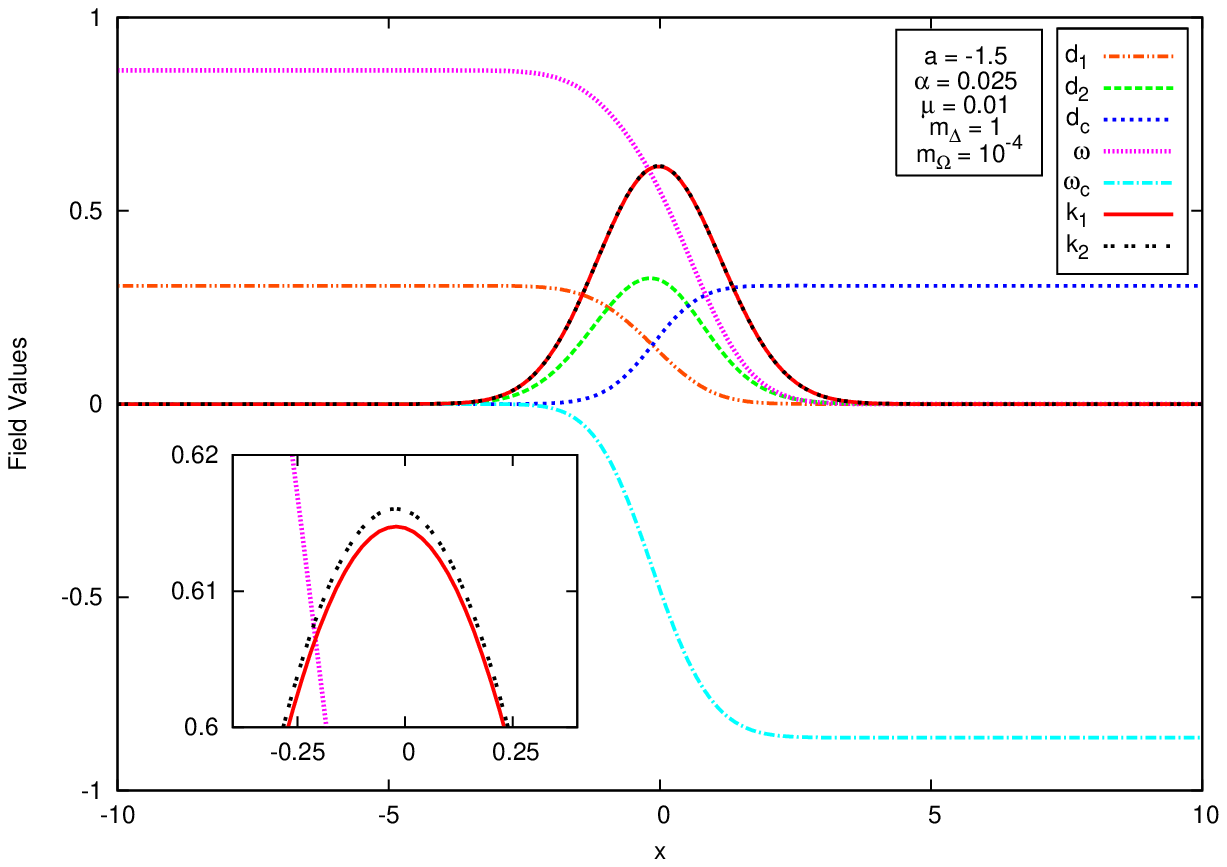}
\caption{Domain wall with $CP$ violating condensate in MSLR\rlap/P. 
The inset magnifies the behaviour of the $k_1$ and $k_2$ near their maximum value.}
\label{fig:panelMSLRPslash}
\end{center}
\end{figure}

\begin{table}[!htp]
\begin{center}
\begin{tabular}{c|c|c|c|r@{.}l}
\hline
\centering{$\alpha$} & \centering{$k_1$} & 
\centering{$k_2$} & $\chi $ & 
\multicolumn{2}{c}{$\chi-\pi/4$} \\
\hline \hline
0.001 & 0.6170 & 0.6176 & 0.7859 &  0&0005 \\
0.005 & 0.6173 & 0.6183 & 0.7861 &  0&0007 \\
0.01  & 0.6173 & 0.6185 & 0.7863 &  0&0009 \\
0.025 & 0.6147 & 0.6160 & 0.7864 &  0&0010 \\
0.035 & 0.6108 & 0.6116 & 0.7860 &  0&0006 \\
0.045 & 0.6055 & 0.6053 & 0.7852 & -0&0001 \\
0.05  & 0.6023 & 0.6015 & 0.7847 & -0&0006 \\ 
0.10  & 0.5572 & 0.5467 & 0.7758 & -0&0095 \\
0.15  & 0.5042 & 0.4815 & 0.7624 & -0&0229 \\
0.20  & 0.4564 & 0.4225 & 0.7468 & -0&0385 \\
0.25  & 0.4178 & 0.3740 & 0.7301 & -0&0552 \\
0.30  & 0.3879 & 0.3354 & 0.7128 & -0&0725 \\
0.50  & 0.3233 & 0.2400 & 0.6386 & -0&1467 \\ 
0.75  & 0.2889 & 0.1745 & 0.5433 & -0&2420 \\
1.00  & 0.2661 & 0.1311 & 0.4579 & -0&3274 \\
 \hline \hline
\end{tabular}
\caption{ Peak phase values $\chi = \mbox{tan}^{-1}(k_2/k_1)$ in both MSLRM and MSLR\rlap/P 
for various values of $\alpha$ 
}
\label{tab:peakchi}
\end{center} 
\end{table} 

\begin{equation} 
\begin{array}{ccc}
\langle \Omega \rangle = 
\begin{pmatrix}
\omega & 0 \\
0 & - \omega
\end{pmatrix}, & \qquad
\langle \Delta \rangle =
\begin{pmatrix}
0 & 0 \\
d_1 + id_2 & 0
\end{pmatrix}, & \qquad
\langle \bar{\Delta} \rangle =
\begin{pmatrix}
0 & \sqrt{d_1^2 + d_2^2} \\
0 & 0
\end{pmatrix}, \\[0.75cm]
\langle \Omega_c \rangle = 
\begin{pmatrix}
\omega_c & 0 \\
0 & - \omega_c
\end{pmatrix}, & \qquad
\langle \Delta_c \rangle =
\begin{pmatrix}
0 & 0 \\
d_c & 0
\end{pmatrix}, & \qquad
\langle \bar{\Delta}_c \rangle =
\begin{pmatrix}
0 & d_c \\
0 & 0
\end{pmatrix}.
\label{eq:simvev} 
\end{array}
\end{equation}
\begin{equation}
\begin{array}{cc}
\langle \Phi_1 \rangle =
\begin{pmatrix}
k_1 + ik_2 & 0 \\
0 & \sqrt{k_1^2 + k_2^2}
\end{pmatrix},
& \qquad
\langle \Phi_2 \rangle =
\begin{pmatrix}
\sqrt{k_1^2 + k_2^2} & 0 \\
0 & \sqrt{k_1^2 + k_2^2}
\end{pmatrix}
\label{eq:simphi} 
\end{array}
\end{equation} 
The effective potential obtained by substituting these vev's is given in 
eq. (\ref{eq:simpot}) in the appendix \ref{sec:simpVeff}. 
In accordance with the discussion accompanying eq.s (\ref{eq:vevvalues}) and
(\ref{eq:scale-rel}), we choose the scale of 
of $M_{B-L} \sim 10^4$ GeV which relates to $M_R$ being of the order of
$10^6$ GeV. 

For numerical simulation the mass parameters are scaled by the largest scale 
$M_R \sim 10^6$ GeV,  i.e. in our simulation $M_R \sim 1$, and other parameters
are chosen $m_\Delta \sim \mathcal{O}(1)$ and 
$m_\Omega \sim \mathcal{O}(10^{-4})$ as per eq. (\ref{eq:vevvalues}).
Parameter $\mu$ entering the bidoublet mass terms should be $10^{-4}$, however
at the scale in question, due to temperature corrections it is expected
to be of the same order as $M_{B-L}$ and is chosen $0.01$. 
Eq. (\ref{eq:vevvalues}) dictates that the parameter $a$ be negative and
order unity. It is chosen to be $-1.5$ throughout. 
The asymptotic values of the fields are such as to minimize the potential under
translation invariance. The profiles are then found by relaxation methods.
Two examples of the numerically determined profiles are shown in figures
\ref{fig:panelMSLRM} and \ref{fig:panelMSLRPslash}.

Electroweak symmetry is unbroken at the epoch under consideration and
hence the asymptotic values for $k_1$ and $k_2$ are zero. 
Since both $k_1$ and $k_2$ approach the same values asymptotically, the
effective asymptotic value of $\chi$ is $\pi/4$. The departure from
this value at the maxima of the graphs are listed in table \ref{tab:peakchi}.
It was observed that the difference in $k_1$ and $k_2$ profiles, the 
source of spatially varying $CP$ violating phase $\chi$ arises from the terms 
\begin{equation}
 16\,{\mu }^2\, {k_1}\,{\sqrt{k_1^2 + k_2^2}}
 + ~2\,a\,\alpha \,d_c^2\,{k_1}\,{\sqrt{k_1^2 + k_2^2}}
 + ~4 \alpha m_\Omega (\omega - \omega_c) k_1 \sqrt{k_1^2 + k_2^2} .
\end{equation} 
The parameter $\alpha$ entering the superpotential  is the  least 
controlled by the fundamental symmetries and phenomenological 
considerations, and plays a very significant role. Small values 
of $\alpha$ make the difference between $k_1$ and $k_2$ indistinguishable 
in the graphs. Since the final baryon  symmetry after conversion from the 
lepton asymmetry is a small number, such parameter ranges are also of relevance. 
Mid-range values of $\alpha$ are  favorable to make the phase of 
$\chi = \mbox{tan}^{-1}(k_2/k_1)$ more pronounced as can be seen from 
table \ref{tab:peakchi}.

We see in table \ref{tab:peakchi} that the $CP$ phase values in both 
models are identical, other parameters remaining the same. This can be 
seen from the effective 
potential for MSLR\rlap/P worked out in the appendix \ref{sec:simpVeff}. 
The corresponding expression
for the effective potential for MSLRM can be obtained by simply reversing 
the sign of $\omega_c$. However, upon minimizing, the vev for $\omega_c$
also has opposite signs in the two models and hence the $k_1$, $k_2$ see
the same effective potential in the two cases. 

\section{Conclusion}
\label{sec:conclusion}

We have explored two possible realizations of supersymmetric Left-Right 
symmetric model for their implications to cosmology. 
The superpotential imposes the requirements that $SU(2)_R$
breaks first to $U(1)_R$ at a scale $M_R$ and $U(1)_{B-L}$ breaks at
a lower scale $M_{B-L}$ with a see-saw requirement $M_{B-L}^2 \simeq M_R M_{EW}$
with respect to the SM scale $M_{EW}$. This makes it interesting to explore
the values $10^4$ GeV for $B-L$ breaking scale and  $10^6$ GeV for the
$SU(2)_R$ breaking scale. 

The first stage of symmetry breaking makes only local choices of the new
phase, leading to formation of DW, 
which remain metastable down to $M_{EW}$ temperature scale in MSLRM but
only upto a higher scale $m_{B-L}$ in MSLR\rlap/P. After the DW
are rendered metastable. they remain a dominant source of energy
down to a temperature $T_D$ which would depend on the details of
DW evolution dynamics.
Only when the DW have disappeared is the phase transition completed, 
ensuring a unique global choice of chirality. These facts, summarized in
table \ref{tab:pattern} play a central 
role in constraining the models since the DW dynamics is meant to achieve
two important cosmological goals, that of removing unwanted relics by 
inducing secondary or weak inflation and causing leptogenesis. Cosmologically
acceptable values of $T_D$ are shown to constrain soft parameters in the 
Higgs sectors of the two models in table \ref{tab:DWalls} and eq. 
(\ref{eq:a_diff}).
We have presented the explicit solutions
for the DW configurations for a range of parameters and determined the 
possibility of a transient $CP$ violating  phase in the core of the DW.
It is interesting that due to the nature of the effective potential, 
the CP violating phase is quantitatively identical in the two variants
for the same values of the parameters. This is discussed in sec. 
\ref{sec:wallprofiles}.

The MSLRM permits a long duration of cosmological domination by DW.
The disappearance of the DW and the completion of the phase transition is 
signaled only after TeV scale supersymmetry 
breaking. This permits removal of cosmological relics, but also potentially
leptogenesis from the uni-directional motion of the DW. The phase transition 
is expected to end with reheating to a scale above the electroweak scale,
so that thermal leptogenesis mechanism through resonant leptogenesis, 
arising from soft supersymmetry breaking terms is also possible.
It is interesting that the estimate $B\sim 0.1$eV in the leptonic sector
required from thermal leptogenesis is in concordance with the independent 
cosmological requirement on soft parameters in the Higgs sector for
the successful disappearance of the DW.

A new model MSLR\rlap/P has been proposed  for making
global parity breakdown to a unique vacuum natural. It relies on
choosing a phase $-1$ for the $SU(2)$ triplets $\Omega$ 
and $\Omega_c$ under the parity $L\leftrightarrow R$. The first order
phase transition leading to unique global vacuum is signaled in this
model at the higher scale $M_{B-L}$ compared to the case of MSLRM. 
Successful completion of the phase transition in this model also 
relies on the supersymmetry breaking mechanism but it is possible to
impose the stricter requirement that the soft terms obey the gauge and 
discrete symmetries of the superpotential. The uniqueness of the
global vacuum then follows from the spontaneous symmetry breaking 
within the visible sector. Again, as in the MSLRM, resonant soft 
leptogenesis as well as DW mediated leptogenesis remain viable.

There are general arguments based on intrinsic reasons suggesting that 
TeV scale leptogenesis if true cannot be verified in colliders in the 
near future \cite{Hambye:2001eu}. 
We have adopted the approach of \cite{Chun:2005tg} 
wherein  cosmology requirements arising from soft resonant leptogenesis 
are correlated with collider observables. Furthermore, the occurrence of 
a phase transition  accompanied by domain walls may be verifiable in 
upcoming  and planned gravitational wave experiments \cite{Grojean:2006bp}.
An open question for this class of models is
a comprehensive analysis of the two different potential sources 
of leptogenesis,  from phase transition DW and from the resonant 
thermal mechanism. Successful cumulative leptogenesis and subsequent 
dilution to required baryon asymmetry  can further constrain the 
parameters of the models.

\section{Acknowledgment}
\label{sec:acknowledge} 
This work is supported by a grant from the Department of Science and
Technology, India. The work of AS is supported by Council of Scientific 
and Industrial Research, India.

\appendix
\section{$F$ and $D$ Flatness Conditions}
\label{sec:flatness}

The $F$-flatness conditions for MSLR\rlap/P are
\begin{eqnarray}
F_{\bar\Delta} &=& m_\Delta \Delta 
	+ a (\Delta \Omega - \frac{1}{2} {\rm Tr}\,\Delta\Omega) =0 
\nonumber  \\
F_{\bar{\Delta}_c} &=& m_\Delta  \Delta_c  
	- a (\Delta_c \Omega_c - \frac{1}{2} {\rm Tr}\,\Delta_c\Omega_c)=0
\nonumber  \\ 
F_\Delta &=& m_\Delta  \bar\Delta 
	+ a (\Omega\bar\Delta- \frac{1}{2} {\rm Tr}\,\Omega\bar\Delta)=0
\nonumber  \\
F_{\Delta_c} &=& m_\Delta \bar\Delta_c 
	- a (\Omega_c{\bar\Delta_c}-\frac{1}{2}{\rm Tr}\,
	\Omega_c\bar\Delta_c) =0
\nonumber  \\
F_\Omega &=& m_\Omega \Omega
	+ a (\bar\Delta\Delta  - \frac{1}{2} {\rm Tr}\,\bar\Delta\Delta) 
	+ \alpha_{ij} \tau^T_2 \Phi_j \tau^T_2 \Phi^T_i = 0 
\nonumber \\    
F_{\Omega_c} &=& m_{\Omega} \Omega_c
   - a (\bar\Delta_c\Delta_c  - \frac{1}{2} {\rm Tr}\,\bar\Delta_c
   \Delta_c) - \alpha_{ij} \tau^T_2 \Phi^T_j \tau^T_2 \Phi_i = 0
\nonumber \\
F_{\Phi_i} &=& \alpha_{ij}( \Omega^T
    \tau_2^T \Phi_j \tau_2^T - \tau_2 \Omega \Phi_j \tau_2 
    - \tau_2 \Phi_j \tau_2 \Omega_c + \tau_2^T \Phi_j
    \Omega_c^T \tau_2^T) 
\nonumber \\ && 
    + ~\mu_{ij} (\tau_2^T \Phi_j \tau_2^T
    + \tau_2 \Phi_j \tau_2) = 0 
\label{eq:f-flat}
\end{eqnarray}
The $D$-flatness conditions for MSLR\rlap/P are given by
\begin{eqnarray}
D_{R i} &= & 2 {\rm Tr}\,\Delta_c^\dagger\tau_i\Delta_c + 2 
{\rm Tr}\,\bar\Delta_c^\dagger\tau_i\bar \Delta_c 
+ 2 {\rm Tr}\,\Omega_c^\dagger\tau_i\Omega_c
= 0 
\nonumber \\
D_{L i} &=&  2 {\rm Tr}\,\Delta^{ \dagger}\tau_i\Delta 
+ 2 {\rm Tr}\,\bar\Delta^{ \dagger}\tau_i\bar \Delta 
+  2 {\rm Tr}\,\Omega^{ \dagger}\tau_i\Omega 
= 0 
\nonumber  \\
D_{B-L} &=& 
2 {\rm Tr}\,(\Delta^{ \dagger}\Delta 
- \bar\Delta^{ \dagger}\bar \Delta)
- 2 {\rm Tr}\,(\Delta_c^\dagger\Delta_c 
- \bar\Delta_c^\dagger\bar \Delta_c )=0
\label{eq:d-flat}
\end{eqnarray}
Since the Leptons $L$ and $L_c$ are considered to have zero vev, we omit
them from the $F$ and $D$ flat conditions.
The above conditions are same for MSLRM with only $\Omega_c$ replaced 
by $-\Omega_c$. 

\section{Simplified effective potential}
\label{sec:simpVeff}
Here we display the simplified effective potential involving seven 
degrees of freedom referred to in sec. \ref{sec:wallprofiles}.

\begin{eqnarray}
V_{7dof} &=&
\frac{a^2}{2} \left( \left(d_1^2 + d_2^2 \right)^2 + d_c^4 \right) 
+ 2\,a^2\, \left( \omega^2\, \left(d_1^2 + d_2^2 \right) 
+ \omega_c^2 d_c^2 \right)
\nonumber \\ &&
+ 16\,{\mu }^2\, ( 3 (k_1^2 + k_2^2)
+ {k_1}\,{\sqrt{k_1^2 + k_2^2}} )
+ 16\,\alpha^2\, (\omega - \omega_c)^2 (k_1^2 + k_2^2)
\nonumber \\ &&
+ 8\,\alpha^2 (k_1^2 + k_2^2)^2
- 8 \alpha^2 k_1 (k_1^2 + k_2^2)^{3/2} 
- 2a\alpha d_1 {\sqrt{d_1^2 + d_2^2}} (k_1^2 + k_2^2)
\nonumber \\ &&
+ 2\,a\,\alpha (d_1 k_1 + d_2 k_2) {\sqrt{d_1^2 + d_2^2}}\,
{\sqrt{k_1^2 + k_2^2}} + 
2\,a\,\alpha \,d_c^2\,{k_1}\,{\sqrt{k_1^2 + k_2^2}} 
\nonumber \\ &&
- 2\,a\,\alpha \,d_c^2\, (k_1^2 + k_2^2)
+ 4\,a\,m_\Delta \left(\omega (d_1^2 + d_2^2) -\omega_c d_c^2 \right)
\nonumber \\ &&
+ 2 m_\Delta^2 (d_1^2 + d_2^2 + d_c^2)
+ 2\,a\,m_\Omega \left( \omega \,d_1\,{\sqrt{d_1^2 + d_2^2}}\,
- \omega_c d_c^2 \right)
\nonumber \\ &&
- 4 \alpha m_\Omega (\omega - \omega_c) \left( k_1^2 + k_2^2 
- k_1 \sqrt{k_1^2 + k_2^2} \right)
+ 2\,m_\Omega^2 (\omega^2+\omega_c^2)
\label{eq:simpot} 
\end{eqnarray}






\end{document}